%% file: IPN_S56VSR123grb_prl.tex
\documentclass[prl,twocolumn,lengthcheck,superscriptaddress,showpacs,letterpaper,nofootinbib]{revtex4-1}

\usepackage[utf8]{inputenc}
\usepackage[T1]{fontenc}
\usepackage{ae,aecompl} 
\usepackage{graphicx}
\usepackage{amsmath}
\usepackage{color}
\usepackage{amssymb}
\usepackage{latexsym}
\usepackage{wasysym}
\usepackage{psfrag}
\usepackage{comment}
\usepackage{ifthen}
\usepackage[colorlinks=true]{hyperref}
\usepackage{longtable}
\usepackage{float}
\restylefloat{table}

\usepackage{natbib}
\def\be{\begin{equation}}
\def\ee{\end{equation}}
\def\ben{$$}
\def\een{$$}
\def\bea{\begin{eqnarray}}
\def\eea{\end{eqnarray}}
\def\bean{\begin{eqnarray*}}
\def\eean{\end{eqnarray*}}

\def\bi{\begin{itemize}}
\def\ei{\end{itemize}}
\def\ben{\begin{enumerate}}
\def\een{\end{enumerate}}

\newcommand{\makevisible}[1]{\textbf{#1}}
\newcommand{\switch}[1]{%
  \ifthenelse{\equal{#1}{0}}{\renewcommand{\makevisible}[1]{}}{}}
\switch{0} 

\newcommand{\distBurst}{13}
\newcommand{\DistNSNS}{12}
\newcommand{\DistNSBH}{22}
\newcommand{\nAnalyzedGRB}{223}
\newcommand{\nBurstGRB}{221}
\newcommand{\nAnalShort}{27}
\newcommand{\nBurstShort}{25}

\def\apj{Astrophysical Journal }  
\def\apjl{Astrophysical Journal } 

\def\mnras{MNRAS }                
\def\prd{Phys.~Rev.~D }           
\def\nat{Nature }                 
\def\physrep{Phys.~Rep.}

\def\version$#1,v #2 #3${#2}

\newcommand{\dccversion}{10}

\begin{document}
\title{Search for gravitational waves associated with gamma-ray bursts detected by the InterPlanetary Network}
\input{LSC_Feb2014_Virgo_Feb2014-prd}

\begin{abstract}

We present the results of a search for gravitational waves associated with \nAnalyzedGRB\,gamma-ray bursts (GRBs) detected by the InterPlanetary Network (IPN) in 2005--2010 during LIGO's fifth and sixth science runs and Virgo's first, second and third science runs. The IPN satellites provide accurate times of the bursts and sky localizations that vary significantly from degree scale to hundreds of square degrees.  We search for both a well--modeled binary coalescence signal, the favored progenitor model for short GRBs, and for generic, unmodeled gravitational wave bursts. Both searches use the event time and sky localization to improve the gravitational-wave search sensitivity as compared to corresponding all--time, all--sky searches. We find no evidence of a gravitational-wave signal associated with any of the IPN GRBs in the sample, nor do we find evidence for a population of weak gravitational-wave signals associated with the GRBs. For all IPN--detected GRBs, for which a sufficient duration of quality gravitational-wave data is available, we place lower bounds on the distance to the source in accordance with an optimistic assumption of gravitational-wave emission energy of $10^{-2}M_{\odot}c^2$ at 150 Hz, and find a median of \distBurst\,Mpc. For the \nAnalShort\ short-hard GRBs we place 90\% confidence exclusion distances to two source models: a binary neutron star coalescence, with a median distance of \DistNSNS\,Mpc, or the coalescence of a neutron star and black hole, with a median distance of \DistNSBH\,Mpc. Finally, we combine this search with previously published results to provide a population statement for GRB searches in first--generation LIGO and Virgo gravitational-wave detectors, and a resulting examination of prospects for the advanced gravitational--wave detectors.
\end{abstract}



\maketitle


Gamma-ray bursts (GRBs) are amongst the most energetic electromagnetic astrophysical phenomena. They fall into two commonly accepted groups depending on their duration and spectral hardness \citep{Nakar:2007, Qinx:2010kp} and are referred to as ``short'' or ``long''. 
Short GRBs (duration less than 2s; hard spectra) are believed to be produced by the mergers of either double neutron star or neutron star-black hole binaries \citep{Nakar:2007} and the recent observation of a kilonova associated with GRB130603B \citep{Tanvir:2013,Berger:2013} lends support to this hypothesis. Such compact binary coalescences generate strong gravitational waves (GWs) in the sensitive frequency band of Earth-based gravitational-wave detectors \citep{Blanchet:2001aw,BD89}. The detection of gravitational waves associated with a short GRB would provide direct evidence that the progenitor is indeed a coalescing compact binary as well as distinguish between a double neutron star (NSNS) and neutron star-black hole (NSBH) progenitor.
Long GRBs (duration greater than 2s; soft spectra) models are mostly related to the collapse of rapidly rotating massive stars. The extreme conditions encountered in such objects may make the system susceptible to a variety of rotational instabilities that may emit up to $10^{-2} M_{\odot}c^{2}$ through GW radiation \citep{Modjaz:2011, HjorthBloom:2011}.

Between 2005 and 2010, the first generation of large scale interferometric gravitational-wave detectors, LIGO, Virgo and GEO, were operating at, or close to, their design sensitivities. During these runs, the detectors had a sensitivity out to tens of Mpc for binary mergers \citep{Abbott:2009tt, Abbott:2009qj, Abadie:2010yb, Colaboration:2011np} and other transients emitting $10^{-2}M_{\odot}c^{2}$ in gravitational waves \citep{Abbott:2009zi, Abadie:2010mt, Abadie:2012rq}.
 
Although it is expected that most GRB progenitors observed by gamma-ray detectors will be at distances too large for the resulting gravitational-wave signals to be detectable by initial LIGO and Virgo \citep{berger05, Metzger:2011bv}, it is  possible that in the GRB data set under study that one might be within the range of the detectors. For example, the smallest observed redshift to date of an optical GRB afterglow is $z=0.0085$ ($\simeq 36$\,Mpc) for GRB~980425 \citep{kulkarni98,galama98,iwamoto98}; this would be within the LIGO-Virgo detectable range for some progenitor models. Although GRB~980425 is a long duration soft spectrum GRB, observations seem to suggest that, on average, short--duration GRBs tend to have smaller redshifts than long GRBs \citep{GuPi:05,fox05}.  We therefore search for evidence of a gravitational wave signal associated to any observed short or long GRB for which there is a sufficient duration of high quality data in at least two detectors.
By making use of the known time and sky location of the observed GRB, it is possible to significantly reduce the parameters of the search and consequently improve the sensitivity over an all--sky all--time search of the data. 
Several searches for gravitational waves associated with gamma-ray bursts have been performed in the past using data from both LIGO and Virgo detectors \citep{abbottgrb05,burstGrbS234,Ac_etal:07,Ac_etal:08,lvc_longGRB2013}. Indeed, the data from LIGO's fifth and sixth science runs (S5 and S6) and Virgo's first through third science runs (VSR1-3) were analyzed to search for gravitational wave signals associated with both short and long GRBs observed with the Swift BAT and Fermi GBM and LAT detectors \citep{Abbott:2009kk, Abadie:2010uf, lvc:s6grb}. No evidence for a gravitational wave signal was found in these searches. 

The InterPlanetary Network (IPN) \citep{Hurley:2002wv}, is a group of satellites orbiting the Earth, Mars and Mercury and operating, among other equipment on board, gamma-ray detectors. The IPN, in its current configuration, acts as a quasi--all--sky and full--time gamma-ray burst detector.  Thus, the IPN provides an additional population of GRBs which may not be observed solely by Swift or Fermi and tends to detect brighter (therefore, on average, closer) bursts, which are relevant for gravitational wave searches.
The IPN provides accurate GRB times as well as sky locations determined by triangulation between the satellites in the network. Depending upon the satellites involved, the localization can vary from less than one square degree to over one thousand. Two short duration GRBs, GRB~051103 \citep{Frederiks:2007,Hurley2010} and GRB~070201 \citep{Mazets:2008},  were localized by the IPN with error boxes overlapping the M81 galaxy at 3.6\,Mpc and the Andromeda galaxy (M31) at 770\,kpc, respectively.  The gravitational wave data around the times of these GRBs were analyzed in \cite{Abadie:2012bz} and \cite{Abbott:2007rh}, respectively. The non--detection of associated gravitational waves ruled out the progenitor object being a binary merger in M81 or M31 with over 99\% confidence.

In this paper we present the results of a targeted search for gravitational waves around the burst trigger times of \nAnalyzedGRB\ additional gamma-ray bursts, including \nAnalShort\ short GRBs, localized by the InterPlanetary Network (IPN) during both LIGO's fifth and Virgo's first science runs (S5/VSR1) and LIGO's sixth and Virgo's second and third science runs (S6/VSR2-3). 
 The search for gravitational wave bursts (GWBs) is performed on all the GRBs, short or long, for which we have good--quality gravitational wave data, regardless of the localization error box size. In addition, a search for a binary merger signal is performed for the \nAnalShort\ short hard GRBs for which there was sufficiently good sky localization to make the search tractable.

We find no evidence for a gravitational wave candidate associated with any of the IPN GRBs in this sample, and statistical analyses of the GRB sample rule out the presence of a collective signature of weak gravitational waves associated with the GRB population. We place lower bounds on the distance to the progenitor for each GRB, and constrain the fraction of the observed GRB population at low redshifts.  Additionally, we combine the results presented here with those from previous analyses to provide a comprehensive limit from all of the GRB searches.  Using this, we extrapolate to the future advanced detector network and show that the observation of a gravitational wave signal associated with a GRB is possible, but by no means guaranteed and the continued all sky coverage provided by the IPN will increase these chances.


\section{GRB sample}
\label{sec:grbsample}

The Interplanetary Network (IPN) is a group of spacecraft equipped with gamma-ray detectors used to localize gamma-ray bursts (GRBs) and soft gamma repeaters (SGRs, or magnetars). The IPN has contributed burst data to LIGO since the initial engineering runs in 2001. At the time of this combined search, nine spacecraft contribute their data: Wind, Mars Odyssey, MESSENGER, INTEGRAL, RHESSI, Swift, Suzaku, AGILE and Fermi. 

The astronomical locations of GRBs are determined by comparing the relative arrival times of the signal at the spacecraft. The precision is inversely proportional to the spacecraft separations, among other things, so that the localization accuracy of a network with baselines of thousands of light-seconds can be equal to or superior to that of any other technique.  A description of the error box construction process can be found in \cite{Hurley:1999ym} and, specific to our search, in \cite{Predoi:2011aa}. The light curves, energy spectra, and localizations of all the bursts in our sample were examined to eliminate the possibility of contamination by magnetar bursts or solar flares. None of these events have been followed--up by X--ray or optical telescopes so no information on afterglows or possible host galaxies and associated redshifts is available. The full list of GRBs and their parameters can be found in the supplementary material associated to the paper.

The classification of GRBs into long or short typically uses the $T_{90}$ duration for a given detector.  This is defined as the time interval over which 90\% of the total background-subtracted photon counts are accumulated.  However this time depends on the energy range the detector is sensitive to, and may therefore vary across the satellites observing the bursts.  Since the IPN bursts are observed by a set of different detectors with different sensitivities, to quote a single $T_{90}$ for them could be misleading.  Even when a single detector measures this time, it is possible to get different numbers for the same burst depending on the arrival angle, which affects the sensitivity as a function of energy.  In this analysis, where possible, we have used the classification provided by \cite{Pal'shin:2013ps}, based on observations with Konus--Wind.  We note that the set of short bursts observed by Konus is split into type I (likely merger scenario), possibly with extended emission, and type II (collapsar). For the modeled search, we only analyzed the type I bursts. For bursts not observed by Konus, the $T_{90}$ observed by Suzaku was used and in cases where this was not available, a by--eye estimate of duration from another mission with good sensitivity (such as Swift or INTEGRAL) was used.  In these cases, any burst with a $T_{90}$ under two seconds was classified as short. The unmodeled search uses many minutes of data around the given time of the burst whereas the modeled search uses a small six--second window around ``short'' bursts to search for the binary coalescence, due to the predicted time difference between merger and GRB emission. It is therefore important that the time of arrival at the Earth be calculated as accurately as possible for these bursts.  When the burst is observed by a near--Earth satellite, this is straightforward and approximated to the spacecraft time.  However, for some poorly localized GRBs only observed by distant satellites, the uncertainty in the Earth--crossing time can be up to approximately five seconds. 

Only the GRBs that occurred when two or more of the LIGO and Virgo detectors were operating in a stable configuration are analyzed. Gravitational-wave data segments that are flagged as being of poor quality are excluded from the analysis.  Thus, although the IPN observed over 600 bursts during the period of interest, only \nAnalyzedGRB\ could be analyzed, of which \nAnalShort\ are classified as short. This nevertheless constitutes the largest GRB sample used to date in such a study.

\section{Gravitational wave detectors}\label{sec:detectors}
In this paper, we discuss results obtained from analyzing data collected by the initial LIGO and Virgo gravitational wave detectors. There were three LIGO detectors: a 4\,km and a 2\,km detector, both at Hanford, WA (referred to as H1 and H2, respectively) and another 4\,km detector in Livingston, LA (L1). The Virgo detector is a 3\,km detector located in Cascina, Italy. Details of these detectors can be found in \cite{S5LIGO} and \cite{VirgoDetector}. From 2005--2010, these detectors were operating at or near design sensitivity. The fifth LIGO Science Run (S5) took place from November 2005 to August 2007 and the sixth science run (S6) ran from June 2009 to October 2010. The second Hanford detector (H2) was not operational during S6. Virgo operated three distinct science runs during this time: VSR1 ran from May to October 2007, VSR2 ran from July 2009 to January 2010 and VSR3 ran from August to October 2010.

The existence of multiple, widely separated detectors around the globe aids our ability to localize signals in blind, all--sky searches.  Additionally, noise artefacts in the detectors caused by terrestrial disturbances are likely to be uncorrelated between the detectors, aiding the ability to distinguish true signals from the noise background. The Hanford and Livingston detectors are separated by 3000\,km which corresponds to 10\,ms travel-time for light, or gravitational waves. The Hanford and Virgo observatories are separated by 27\,ms and the Livingston and Virgo observatories by 26\,ms. 


\section{Search Methodology}
\label{sec:search}

The methods used to search for binary merger signals and GWBs are largely the same as for the previous analysis described in \cite{lvc:s6grb}.  The one major change is the necessity to search the variably shaped, irregularly sized error regions provided by the IPN localizations.  We begin by describing how that is done and then provide a brief review of the remainder of the analysis details, referring the reader to \cite{lvc:s6grb} for more details.   


\textit{Covering the error boxes---}
The analysis of the gravitational–wave data depends on the assumed sky direction to the GRB, since the data must be time–shifted according to the expected time-of-arrival at each detector. This is also weighted by the response of each GW detector to the assumed sky direction. Each GRB localization error box corresponds to a 3–$\sigma$ region determined by the intersection of the IPN timing annuli, with a construction process described in detail in \cite{Predoi:2011aa}. However, most of these IPN error regions are larger than the directional resolution of the GW detector network. Thus, to maximize the likelihood of finding a gravitational-wave signal associated with the GRB, we perform a discrete search across the entire IPN error box by populating it with a grid of points and repeating the search at each of these points. 
The GW detectors have a timing resolution of $\sim 0.5$\,ms, corresponding to a spatial resolution of a few degrees. Therefore, we chose search point separations of approximately 3.6 degrees when only LIGO interferometers' data were used in the search and 1.8 degrees when the more widely separated Virgo detector was also used. The probability distributions of search points over the error boxes were chosen Gaussian for long bursts (to assure that there are proportionally more points for those positions with larger probabilities of containing a signal) and uniform for short bursts (no assumptions made in terms of signal origin within the error box). Short GRBs which could not be localized to better than a few hundred square degrees were not analyzed in the modeled search due to high computational requirements and the negligible increase in sensitivity rendered by a targeted search.  We use simulated signals to determine the sensitivity of our searches.  We distribute these over the IPN error boxes, with the density of simulations  weighted according to the estimated source position probability distribution.  This assures that there are proportionally more simulations for those positions with larger probabilities of being the true GRB signal location.

During the S5 run, two LIGO detectors, H1 and H2, were operational at the Hanford site.  There are a number of bursts for which the only available gravitational wave data are from these two detectors.  Since the detectors are co-located and co-aligned, it is not possible to distinguish different sky locations based on the observed gravitational wave signal.  In this case, there is no need to perform an explicit search over the error box and, for the binary search, all such GRBs can be analyzed.



\textit{Search for GWs from a compact binary progenitor---}

For the binary merger progenitor model, it is believed that the delay between the merger and the emission of gamma--rays will be small [see discussion in \cite{lvc:s6grb}]. We therefore search for binary coalescences with a merger time between 5\,s prior to the GRB and 1\,s afterwards. This is wide enough to allow for potential precursors, some uncertainties in the emission model and in the arrival time of the electromagnetic signal at the IPN spacecraft, as well as for the differences in sensitivity of the IPN detectors.  In addition, we require a minimum of 40 minutes of data available around the time of the GRB.  This ensures both that the detectors were operating stably at the time as well as providing a set of comparable data which can be used to estimate the background of noise events.

For the short GRBs, the data streams from the operational detectors are combined coherently and searched by matched-filtering against a bank of binary merger gravitational waveforms, as described in \cite{Harry:2010fr}. The gravitational waveform emitted by a binary system depends on the masses and spins of the NS and its companion (either NS or BH), as well as on the distance to the source, its sky position, inclination angle, and the polarization angle of the orbital axis. As described above, we tile the sky region with a fixed set of points to search, and we similarly tile the mass space \citep{OwenSathyaprakash98} to provide sensitivity to binaries with component masses greater than $1 M_\odot$ with an upper limit of $3 M_{\odot}$ for any neutron star and a total mass of $25 M_\odot$.  The remaining parameters are handled by maximizing the likelihood analytically over these dimensions.  In addition to matched filtering, the analysis utilizes a number of signal consistency  tests to reject non-stationary, transient noise ``glitches'' in the GW detectors' data \citep{Harry:2010fr}.  

\textit{Search for gravitational wave bursts---}
The procedure used to search for generic short-duration ($\lesssim 1$\,s) GW bursts follows that used in previous GRB analyses \citep{burstGrbS5,lvc:s6grb}.  All GRBs are treated identically regardless of their classification. We search for a GW event between 600\,s prior to the GRB trigger time and 60\,s after or the $T_{90}$ time, whichever is greater. This timescale allows us to take into account almost all of the possible scenarios for a GW emission associated with the GRB; see \cite{lvc:s6grb} for more details. Since the GWB search requires more data around the GRB than the binary merger search, there are GRBs for which we can perform the merger search but not the search for unmodeled transients. 
The data within a $\pm 1.5$\,hr window around the GRB is used to estimate the background of noise events in the data.
The search for a GWB between 60 and 500 Hz is performed by the \textsc{X-Pipeline} algorithm \citep{Sutton:2009gi,Was2012}. Candidate events are then ranked based on their energy and a number of signal consistency tests are applied to reduce the effect of non-stationary noise seen in the GW detectors.
Events occurring at times of known instrumental problems or environmental disturbances are discarded.


\textit{Significance of results---}
The significance of any candidate gravitational--wave event is estimated by using the data surrounding the GRB time.  Specifically, we divide the surrounding data into a large number of blocks of identical length to the search region and calculate the false alarm probability (FAP), or p-value of the event by counting the fraction of off-source trials with an event louder than the one observed.  Where necessary, we artificially time-shift the data from the detectors by several seconds to generate more off-source trials that can be used to estimate the background.

In addition to a single, significant event, it's possible that the data could contain several weak signals.  In order to test whether this is the case, we use a weighted distribution of observed p-values and see whether it is consistent with the expected, uniform distribution of noise.  We also use a weighted binomial test to more quantitatively assess consistency with the no-signal hypothesis.  This test looks for deviations from the null hypothesis in the 5\% tail of lowest p-values weighted by the prior probability of detection (estimated from the GW search sensitivity).  This combination allows us to give more weight to those GRBs for which the gravitational wave network was most sensitive, which are the ones we are most likely to detect.  The result of the weighted binomial test is compared with the distribution obtained from simulated results with p-values uniformly distributed in [0, 1] to evaluate the population significance.  The test is described in greater detail in the appendix of \cite{lvc:s6grb} and in \cite{Was2012}.



\section{Results}

A search for gravitational waves has been performed for a total of \nAnalyzedGRB\ GRBs.  Of these, \nAnalShort\  were classified as short GRBs with a likely binary coalescence progenitor. The gravitational-wave data at the time of these GRBs has been searched for evidence of the coalescence waveforms.  For \nBurstGRB\,bursts, including \nBurstShort\ of the \nAnalShort\ short bursts\,we have performed a search for generic gravitational wave transients in the IPN error box around the time of the GRB.  A full list of all GRBs analysed is available in the supplementary material, where we provide two tables listing the short and long GRBs analysed in this search.

\begin{figure}[t]
\centering
\includegraphics[width=0.38\textwidth]{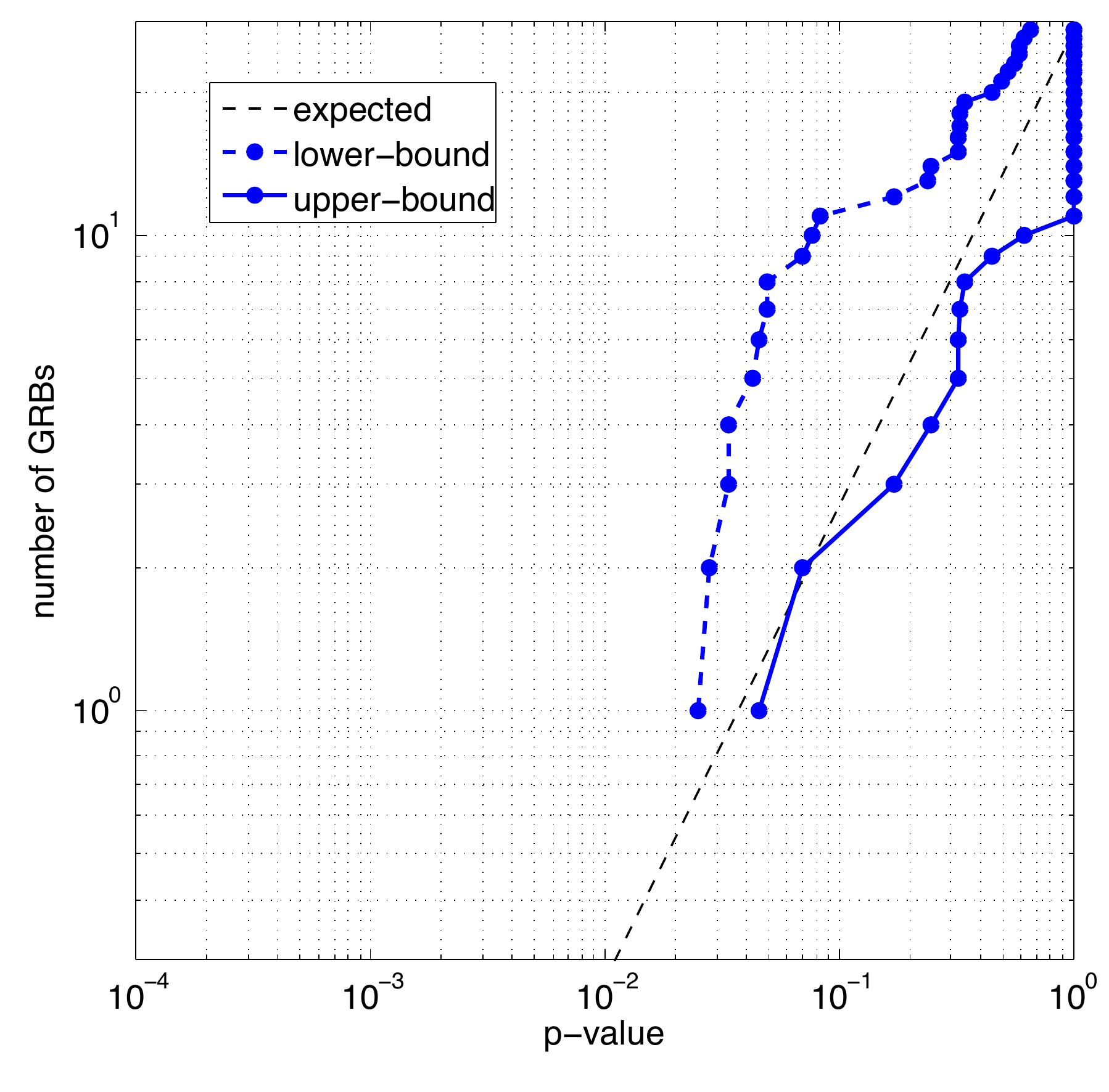}
\caption{Cumulative distribution of p-values from the analysis of 27 short--duration IPN GRBs for evidence of a binary merger gravitational wave signal. The expected distribution under the no--signal hypothesis is indicated by the dashed line. For GRBs with no event in the on--source region, we provide upper bounds on the p-value equal to 1.}
\label{fig:binomialTestCBC}
\end{figure}

\textit{Modeled coalescing binary search results---}
For each GRB, we estimate the p-values of the gravitational wave candidate, by comparing with the background trials.  The distribution of observed p-values is shown in Figure~\ref{fig:binomialTestCBC}.  No significant candidates found.  The p-value is estimated from the background.  However, for a number of GRBs, particularly those observed in the co-located Hanford detectors, the search yields no candidate gravitational wave events after background rejection cuts. For such GRBs, when no event is observed, we cannot quote an exact p-value but only a range bounded below by the fraction of background trials with an event and above by 1.  The result of the weighted binomial population detection test yields a background probability of $\approx$98\%, strongly favouring the no--signal hypothesis. In conclusion, no noteworthy individual events were found by this search, nor evidence for a collective population of weak gravitational wave signals. 

\begin{figure}[t]
\centering
\includegraphics[width=0.38\textwidth]{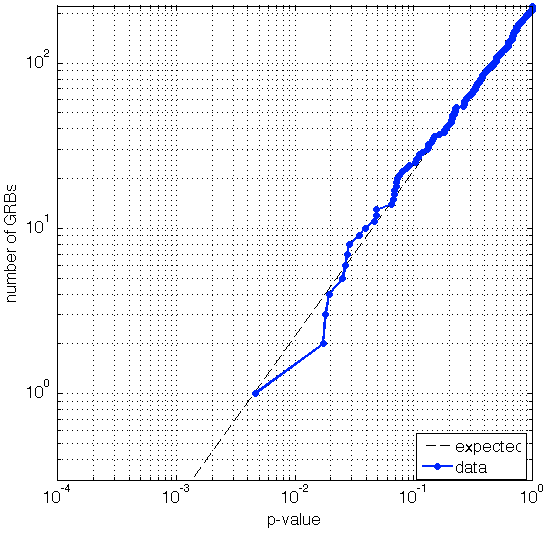}
\caption{Cumulative distribution of p-values from the analysis of 221 IPN GRBs for evidence of a gravitational wave transient associated to the burst. The expected distribution under the no--signal hypothesis is indicated by the dashed line.}
\label{fig:binomialIPNonlyGWB}
\end{figure}

\textit{Unmodeled GW Burst Results---}
The unmodeled gravitational wave burst pipeline analyzed both short and long  GRBs. The distribution of p-values for each of the \nBurstGRB\ IPN GRBs analyzed is shown in Figure \ref{fig:binomialIPNonlyGWB}. The binomial test yields a background probability of 68\%, consistent with the null hypothesis. The smallest p-value, 0.5\%, came from GRB060203B, which is statistically consistent with a no-signal hypothesis given the number analyzed.


\section{Astrophysical interpretation}
\label{sec:discussion}

Given that no significant event was found in our analyses, we place limits on GW emission based on both binary merger (for short GRBs) and generic gravitational wave burst (for all GRBs) signal models, and assess the potential of a similar search with second-generation gravitational-wave detectors around 2015--2020.

\textit{Distance exclusion---}
For a given signal morphology, the gravitational wave analysis is efficient in recovering signals up to a certain distance limit that depends on the sensitivity of the detectors at the time of the search.  We quote a 90\% confidence level lower limit on the distance to each GRB progenitor: that is, the distance at which we recover 90\% of simulated signals.  The quoted exclusion distances are marginalized over systematic errors that are inherent in this analysis: errors introduced by the mismatch of a true GW signal and the waveforms used in the simulations \citep{Abbott:2009tt} and amplitude errors from the calibration of the detector data \citep{Abadie:2010px}.

\begin{figure}[t]
\centering
\includegraphics[width=0.38\textwidth]{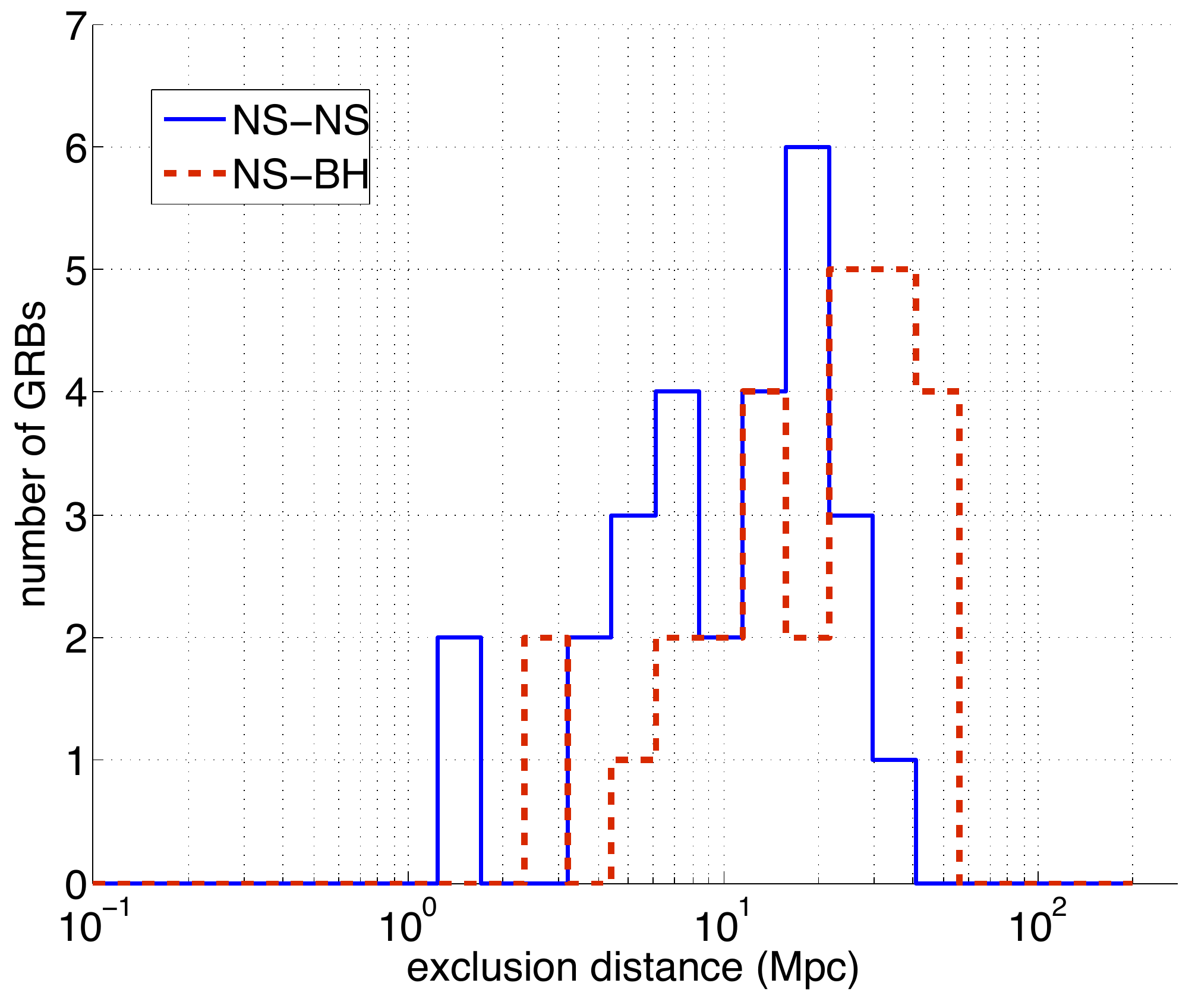}
\caption{Histograms across the sample of short IPN GRBs of the distance exclusions at the 90\% confidence level for NS--NS and NS--BH systems.}
\label{fig:dist_insp}
\end{figure}

For the short GRBs, we calculate a distance exclusion for both two neutron stars (NSNS) and a neutron star with a black hole (NSBH).  In both cases, we assume a jet half--opening angle of 30$^\circ$, and assume that the GRB is emitted in the direction of the binary's total angular momentum.  The median exclusion distance for NSNS is \DistNSNS\ Mpc and for NSBH is \DistNSBH\ Mpc.  A histogram of their values is shown in Figure \ref{fig:dist_insp}.  The NS masses are chosen from a Gaussian distribution centered at 1.4~$\mathrm{M_\odot}$ \citep{Kiziltan:2010ct,Ozel:2012ax} with a width of 0.2~$\mathrm{M_\odot}$ for the NSNS case, and a broader spread of 0.4~$\mathrm{M_\odot}$ for the NSBH systems, to account for larger uncertainties given the lack of observations for such systems. The BH masses are Gaussian distributed with a mean of 10~$\mathrm{M_\odot}$ and a width of~6\,$\mathrm{M_\odot}$. The BH mass is restricted such that the total mass of the system is less than $25\,M_{\odot}$. For masses greater than this distribution, the NS would be swallowed without disruption by the BH, no massive torus would form, and no GRB would be produced \citep{Ferrari:2009bw,Duez:2010,lrr-2011-6}. The dimensionless NS spins are drawn uniformly over [0, 0.4], and the BH spins are drawn uniformly over [0, 0.98) with tilt angle $< 60^{\circ}$.

\begin{figure}[t]
\centering
\includegraphics[width=0.38\textwidth]{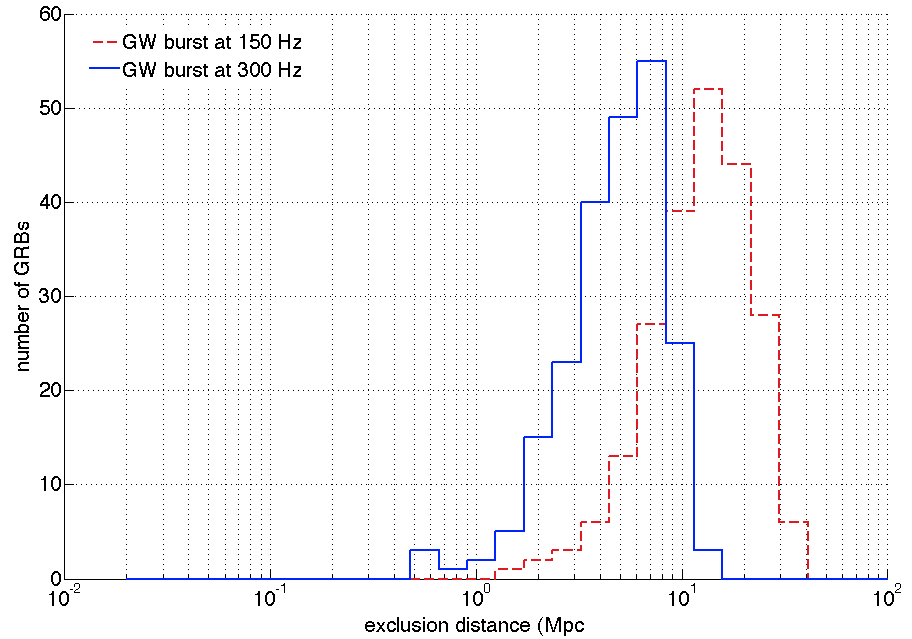}
\caption{Histograms across the sample of IPN GRBs of the distance 
    exclusions at the 90\% confidence level for circularly polarized 
    sine-Gaussian GW burst models at 150\,Hz and 300\,Hz. We assume
    an optimistic standard siren GW emission of 
    $E_\text{GW} = 10^{-2}\,\mathrm{M_\odot c^2}$.}
\label{fig:dist_excl_bursts}
\end{figure}

For the GWB search, no specific waveform model is assumed. Consequently, we use generic signal morphologies to give an idea of the search sensitivity. Specifically, we use circularly polarized sine-Gaussians with central emission frequencies of 150\,Hz and 300\,Hz. We assume a jet opening angle of 5$^\circ$, which is appropriate for long GRBs.  We also assume a total GW emission of 10$^{-2}\,\mathrm{M_\odot c^2}$; this corresponds to the most optimistic models for gravitational wave emission from long GRBs \citep{lvc:s6grb}. The median exclusion distance is 13.0\,Mpc at 150\,Hz and 4.9\,Mpc at 300\,Hz, and histograms of the distributions are given in in Figure \ref{fig:dist_excl_bursts}. 

\begin{figure}[t]
\centering
\includegraphics[width=0.38\textwidth]{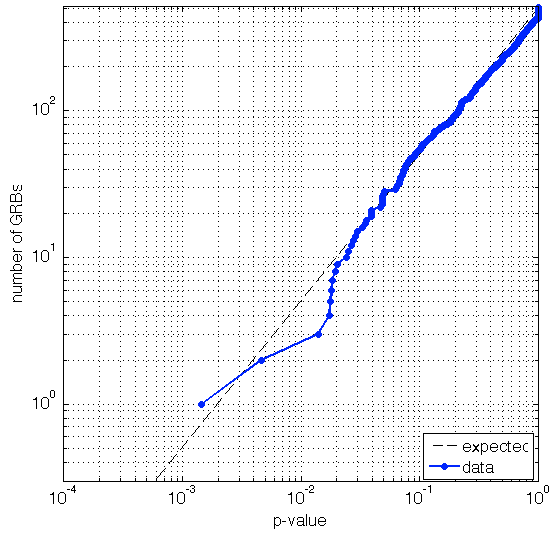}
\caption{Cumulative distribution of p-values from
the analysis of 508 GRBs. The expected distribution under the no--signal hypothesis is indicated by the dashed line.}
\label{fig:binomialallGWB}
\end{figure}

\begin{figure}[t]
  \includegraphics[width=0.38\textwidth]{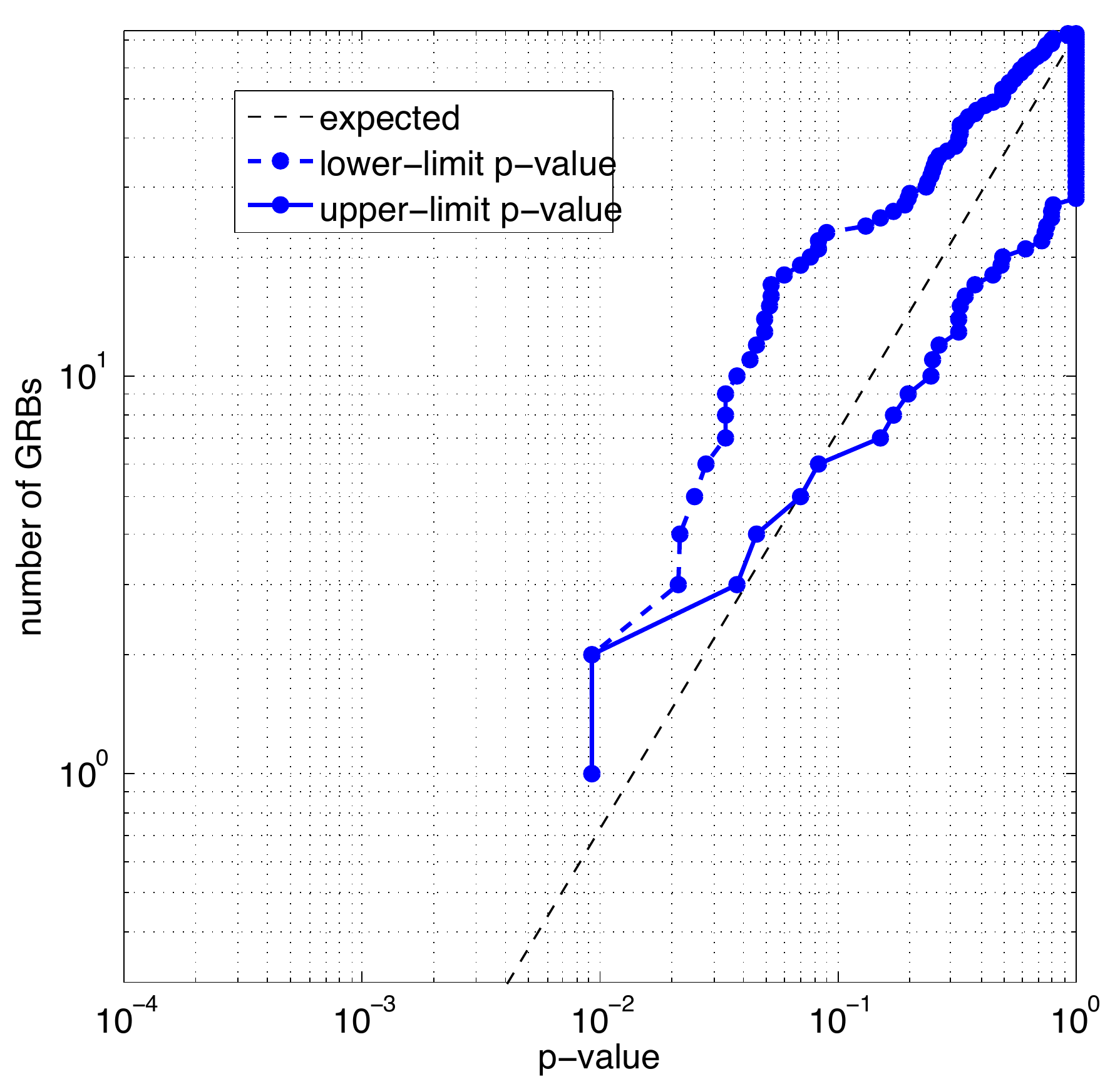}
  \caption{Cumulative distribution of p-values for all 69 analyzed short GRBs. The expected distribution under the no--signal hypothesis is indicated by the dashed line. For GRBs with no event in the on--source region, we provide upper bounds on the p-value of 1.}
  \label{fig:cbc_FAPAll}
\end{figure}

\textit{Population exclusion from all GRBs analyzed---}
Here we present the combination of all S5-6/VSR1-3 searches for coincident GRB/GW signals from this paper, the S5 GRB search \citep{Abbott:2009kk, Abadie:2010uf} and the recent S6/VSR2-3 search \citep{lvc:s6grb}. Algorithms which were used for the first S5 paper were adjusted and reviewed to make sure all analyses were comparable. In total, 508 GRBs were analyzed with the burst search and 69 short GRBs were analyzed for a compact binary coalescence GW signal. 
None of the separate searches showed evidence of a population of weak events. Figures \ref{fig:binomialallGWB} and \ref{fig:cbc_FAPAll} show the distribution of p-values of the full set of GRBs. The weighted binomial test applied to the full population of GRBs confirms that the observed distributions are consistent with the null hypothesis (no signal being observed).

Next, we use the full population of GRBs to place exclusions on GRB populations. To do this, we use a simple population model, where all GRB progenitors have the same GW emission (standard sirens), and perform exclusion on cumulative distance distributions. We parameterize the distance distribution with two components: a fraction F of GRBs distributed with a constant co-moving density rate up to a luminosity distance R, and a fraction 1-F at effectively infinite distance. This simple model yields a parameterization of astrophysical GRB distance distribution models that predict a uniform local rate density and a more complex dependence at redshift >0.1, as the large-redshift part of the distribution is well beyond the sensitivity of current GW detectors. The exclusion is then performed in the (F,R) plane. For details of this method, see Appendix B of \cite{lvc:s6grb}.

\begin{figure}
  \includegraphics[width=0.45\textwidth]{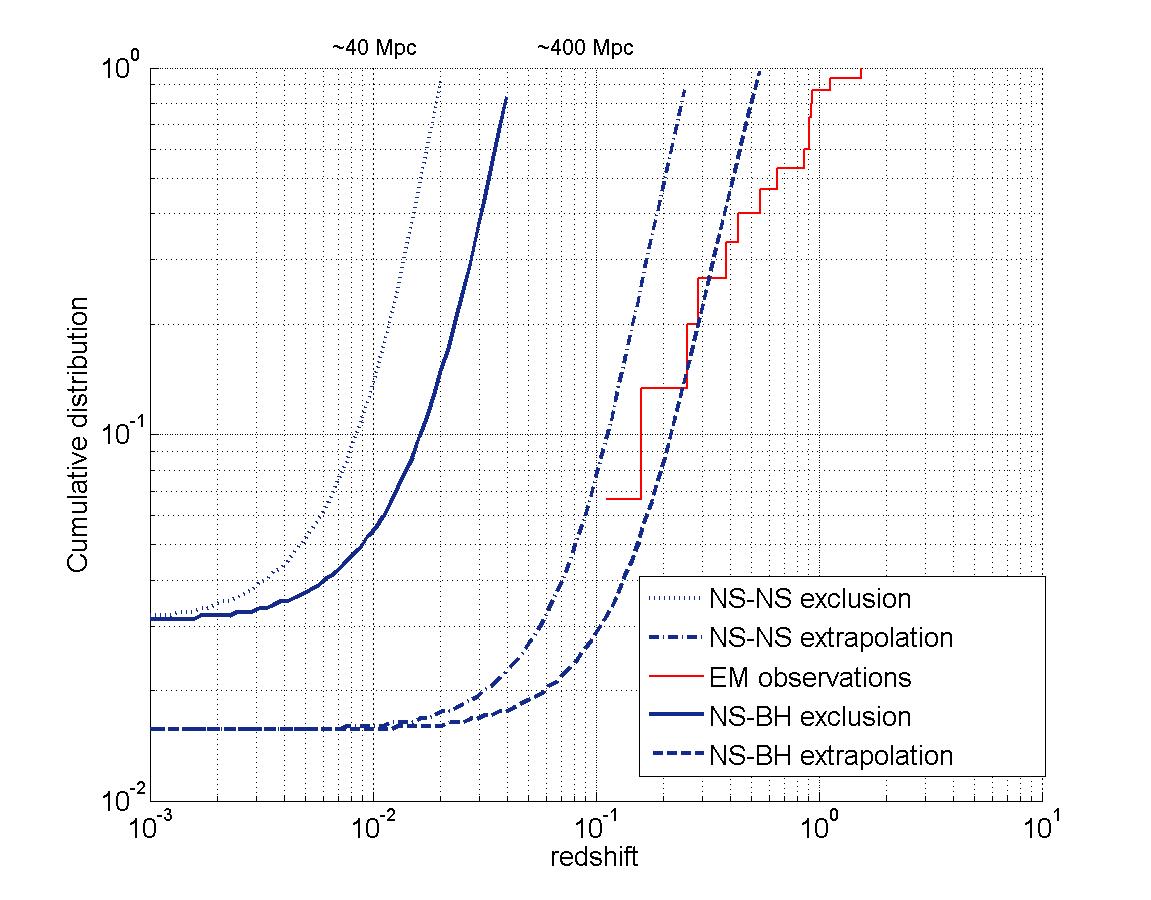}
  \caption{Cumulative distance exclusion for 69 analyzed short GRBs for both a NSNS and a NSBH progenitor model. The exclusion distance is given for this test then extrapolated by a factor of two in number and ten in sensitivity for the advanced detector era expectations. For reference, the red staircase curve shows the cumulative distribution of measured redshifts for \emph{Swift} GRBs~\citep{Jakobsson06,Jakobsson12}.}
  \label{fig:cbc_zExclNSNS}
\end{figure}

In Figure \ref{fig:zExcltot_csg150} we show the exclusion for GW bursts, using as a reference signal a $150$\,Hz sine-Gaussian signal, with an energy in gravitational waves of $10^{-2}\,\mathrm{M_\odot c^2}$.  In addition, we plot the redshift distribution of GRBs as observed by Swift.  
The exclusion at low redshift is dictated by the number of GRBs analyzed and at high redshift by the typical sensitive range of the search. These exclusions assume 100\% purity of the GRB sample.  In Figure \ref{fig:cbc_zExclNSNS}, we show the exclusion for the NS-NS and NS-BH sources.  In neither case does the exclusion line come close to the observed population redshift, indicating that it would have been unlikely to observe an event in this analysis. Indeed, an analysis of all IPN bursts shows that their average redshift is 1.7, and that it detects short bursts with good efficiency up to a redshift of about 0.45.

\begin{figure}
  \includegraphics[width=0.45\textwidth]{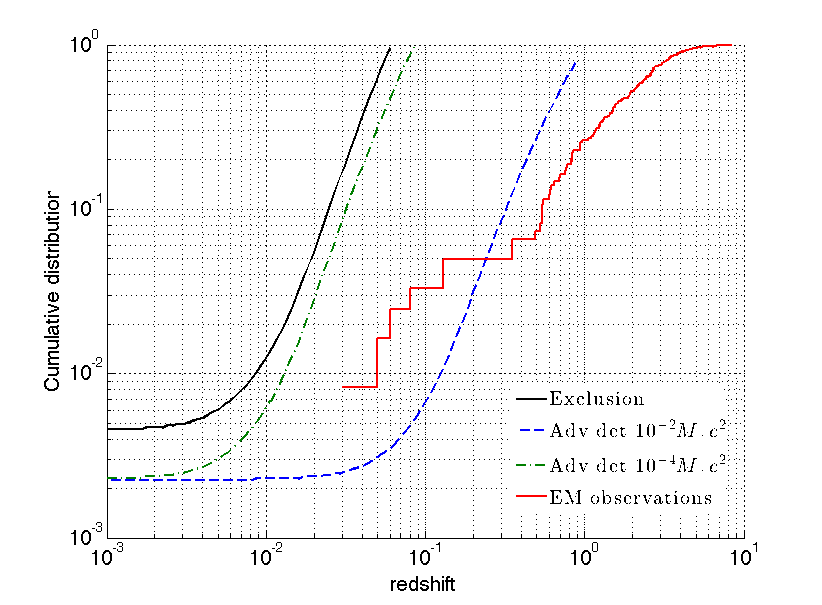}
  \caption{
    Cumulative redshift distribution $F(R)$ exclusion from the analysis of 
    508 GRBs with the GW burst search. We exclude at 90\% confidence level 
    cumulative distance distributions which pass through the region 
    above the black solid curve. We assume a standard siren 
    sine-Gaussian GW burst at $150\,\mathrm{Hz}$ with an
    energy of $E_\text{GW}=10^{-2}\,\mathrm{M_\odot c^2}$.  
    We extrapolate this exclusion to Advanced LIGO/Virgo assuming a
    factor 10 improvement in sensitivity and a factor 2 increase in
    number of GRB triggers analyzed. The blue dashed curve is the
    extrapolation assuming the same standard siren energy of
    $E_\text{GW}=10^{-2}\,\mathrm{M_\odot c^2}$ and the green (gray)
    dashed curve assuming a less optimistic standard siren energy of
    $E_\text{GW}=10^{-4}\,\mathrm{M_\odot c^2}$
    \citep{Ott06,Romero10}. For reference, the red staircase curve
    shows the cumulative distribution of measured redshifts for 
    \emph{Swift} GRBs~\citep{Jakobsson06,Jakobsson12}.
  }
  \label{fig:zExcltot_csg150}
\end{figure}

The advanced LIGO and Virgo detectors are nearing completion. They are expected to start taking data in 2015 and reach their design sensitivity towards the end of the decade \citep{Aasi:2013wya}.  We can use the results obtained here to extrapolate and predict what might be expected with the advanced detectors.  The S5-6 and VSR1-3 runs comprised a total of around 21 months of two (or more) detector duty cycle. Over that period, the detectors' reach varied by approximately a factor of 4, from a 5\,Mpc sensitive distance to NSNS sources for H2 in early S5 to 20 Mpc for H1 and L1 by the end of S6.  Similarly, the current scenario \citep{Aasi:2013wya} calls for around 18 months of science runs of ever increasing sensitivity during the commissioning phase, prior to extended running at design sensitivity which will provide a reach roughly ten times what was achieved in S6 and VSR2-3.  

To approximate the expected advanced detector results, we scale the exclusion distances obtained here by a factor of ten and also increase by a factor of two the number of observed GRBs to account for the increased run time of a few years.  These extrapolated curves are also shown on Figures \ref{fig:zExcltot_csg150} and \ref{fig:cbc_zExclNSNS}.  We see that for both generic burst signals and binary mergers, the exclusion curves are now comparable with the observed redshifts.  In the optimistic scenario where every GRB emits $10^{-2}\mathrm{\,M_\odot c^2}$ of energy in gravitational waves, we will expect to see several such signals, or alternatively be able to exclude this scenario. However, since the $10^{-4}\mathrm{\,M_\odot c^2}$ predicted exclusion is to the left of the observed redshifts, we will be unlikely to observe a gravitational wave associated to a GRB if this is the typical amplitude.  For binary mergers, the NSNS and NSBH extrapolations bracket the line of observed redshifts, indicating that we might expect to see one (or fewer) NSNS associated with a GRB, but might expect several if NSBH are the progenitors.%
\footnote{These extrapolations are broadly comparable to those obtained using only the S6/VSR2-3 \citep{lvc:s6grb}.  They are, however, slightly more pessimistic as they use a more realistic estimate of the evolution of detector sensitivity, as described above.}

Of course, all of the above relies heavily on the continued operation of all sky sensitivity to GRBs, and reasonable localization ability.  Swift and Fermi will continue to run, and SVOM is expected in the advanced detector timeline, but it's clear that the addition of the all--sky, full--time IPN will aid these searches as well.

\section{Conclusion}
\label{sec:conclusion}

We have performed a search for gravitational waves coincident with 223 gamma-ray bursts localized by the InterPlanetary Network over 2005-2010.
These GRBs were detected by the IPN during LIGO’s fifth and sixth science runs and Virgo’s first, second and third science runs.
Of these, we analyzed the 27 short GRBs with a focused search that looked for gravitational-wave signals from the merger of neutron star - neutron star or neutron star - black hole binaries, as most likely expected for short GRBs. 
We also performed an unmodeled burst search over 221 GRBs, both short and long.
No gravitational wave was detected in coincidence with a GRB, and lower limits on the distance were set for each GRB for various gravitational-wave emission models.

Finally, we have combined these results with those of previous analyses for GRBs during S5-6 and VSR1-3 to provide a comprehensive statement on gravitational-wave emission by GRBs from the first-generation LIGO and Virgo detectors. We also extrapolate this exclusion distance for the advanced detector era where we assume a factor of two increase in GRBs and a factor of ten improvement in sensitivity. This shows that the advanced detector era will begin to exclude certain expected models and possibly make coincident gravitational wave detections with GRBs.
These results and prospects for the advanced detector era demonstrate the benefit of searches triggered by gamma-ray burst observatories with high sky--coverage such as the InterPlanetary Network. The continued operation of these satellites will be crucial in future coincident GRB/gravitational wave searches and increase the likelihood of a detection in the advanced detector era.


\acknowledgments
\input{L-V_ack_Jun2013}
KH acknowledges IPN support from the following sources: NASA NNX06AI36G, NNX08AB84G, NNX08AZ85G, NNX09AV61G, and NNX10AR12G (Suzaku); NASA NNG06GI89G, NNX07AJ65G, NNX08AN23G, NNX09AO97G, NNX10AI23G (Swift); NASA NNG06GE69G, NNX07AQ22G, NNX08AC90G, NNX08AX95G, NNX09AR28G (INTEGRAL); The Konus-Wind experiment is partially supported by a Russian Space Agency contract and RFBR grants 12-02-00032-a and 13-02-12017-ofi\_m.
This document has been assigned LIGO Laboratory document number LIGO-P1300226-v\dccversion.

\phantom{aaa}
\bibliographystyle{apsrev4-1}
\bibliography{IPN_Bibliography}


\end{document}

%% file: LSC_Feb2014_Virgo_Feb2014-prd.tex
%
%

\author{%
J.~Aasi$^{1}$,
B.~P.~Abbott$^{1}$,
R.~Abbott$^{1}$,
T.~Abbott$^{2}$,
M.~R.~Abernathy$^{1}$,
F.~Acernese$^{3,4}$,
K.~Ackley$^{5}$,
C.~Adams$^{6}$,
T.~Adams$^{7}$,
P.~Addesso$^{8}$
% ...
}
\collaboration{The LIGO Scientific Collaboration and the Virgo Collaboration}
\noaffiliation

\author{
R.~L.~Aptekar$^{127}$,
J.~L.~Atteia$^{128,129}$,
T.~Cline$^{39}$,
V.~Connaughton$^{130}$,
D.~D.~Frederiks$^{127}$,
S.~V.~Golenetskii$^{127}$,
K.~Hurley$^{131}$,
H.~A.~Krimm$^{132,133}$,
M.~Marisaldi$^{134}$,
V.~D.~Pal'shin$^{127,135}$,
D.~Palmer$^{136}$,
D.~S.~Svinkin$^{127}$,
Y.~Terada$^{137}$,
A.~{von Kienlin}$^{138}$
} \noaffiliation

\affiliation {LIGO, California Institute of Technology, Pasadena, CA 91125, USA }
\affiliation {Louisiana State University, Baton Rouge, LA 70803, USA }
\affiliation {Universit\`a di Salerno, Fisciano, I-84084 Salerno, Italy }
\affiliation {INFN, Sezione di Napoli, Complesso Universitario di Monte S.Angelo, I-80126 Napoli, Italy }
\affiliation {University of Florida, Gainesville, FL 32611, USA }
\affiliation {LIGO Livingston Observatory, Livingston, LA 70754, USA }
\affiliation {Cardiff University, Cardiff, CF24 3AA, United Kingdom }
\affiliation {University of Sannio at Benevento, I-82100 Benevento, Italy, and INFN, Sezione di Napoli, I-80100 Napoli, Italy. }
\affiliation {Albert-Einstein-Institut, Max-Planck-Institut f\"ur Gravitationsphysik, D-30167 Hannover, Germany }
\affiliation {Nikhef, Science Park, 1098 XG Amsterdam, The Netherlands }
\affiliation {LIGO, Massachusetts Institute of Technology, Cambridge, MA 02139, USA }
\affiliation {Instituto Nacional de Pesquisas Espaciais, 12227-010 - S\~{a}o Jos\'{e} dos Campos, SP, Brazil }
\affiliation {International Centre for Theoretical Sciences, Tata Institute of Fundamental Research, Bangalore 560012, India. }
\affiliation {Syracuse University, Syracuse, NY 13244, USA }
\affiliation {University of Wisconsin--Milwaukee, Milwaukee, WI 53201, USA }
\affiliation {Leibniz Universit\"at Hannover, D-30167 Hannover, Germany }
\affiliation {Universit\`a di Siena, I-53100 Siena, Italy }
\affiliation {INFN, Sezione di Pisa, I-56127 Pisa, Italy }
\affiliation {Stanford University, Stanford, CA 94305, USA }
\affiliation {The University of Mississippi, University, MS 38677, USA }
\affiliation {California State University Fullerton, Fullerton, CA 92831, USA }
\affiliation {INFN, Sezione di Roma, I-00185 Roma, Italy }
\affiliation {University of Birmingham, Birmingham, B15 2TT, United Kingdom }
\affiliation {Albert-Einstein-Institut, Max-Planck-Institut f\"ur Gravitationsphysik, D-14476 Golm, Germany }
\affiliation {Montana State University, Bozeman, MT 59717, USA }
\affiliation {European Gravitational Observatory (EGO), I-56021 Cascina, Pisa, Italy }
\affiliation {LIGO Hanford Observatory, Richland, WA 99352, USA }
\affiliation {SUPA, University of Glasgow, Glasgow, G12 8QQ, United Kingdom }
\affiliation {APC, AstroParticule et Cosmologie, Universit\'e Paris Diderot, CNRS/IN2P3, CEA/Irfu, Observatoire de Paris, Sorbonne Paris Cit\'e, 10, rue Alice Domon et L\'eonie Duquet, F-75205 Paris Cedex 13, France }
\affiliation {Columbia University, New York, NY 10027, USA }
\affiliation {Universit\`a di Pisa, I-56127 Pisa, Italy }
\affiliation {CAMK-PAN, 00-716 Warsaw, Poland }
\affiliation {Astronomical Observatory Warsaw University, 00-478 Warsaw, Poland }
\affiliation {Universit\`a degli Studi di Genova, I-16146 Genova, Italy }
\affiliation {INFN, Sezione di Genova, I-16146 Genova, Italy }
\affiliation {Moscow State University, Moscow, 119992, Russia }
\affiliation {LAL, Universit\'e Paris-Sud, IN2P3/CNRS, F-91898 Orsay, France }
\affiliation {NASA/Goddard Space Flight Center, Greenbelt, MD 20771, USA }
\affiliation {University of Western Australia, Crawley, WA 6009, Australia }
\affiliation {Department of Astrophysics/IMAPP, Radboud University Nijmegen, P.O. Box 9010, 6500 GL Nijmegen, The Netherlands }
\affiliation {Universit\'e Nice-Sophia-Antipolis, CNRS, Observatoire de la C\^ote d'Azur, F-06304 Nice, France }
\affiliation {Institut de Physique de Rennes, CNRS, Universit\'e de Rennes 1, F-35042 Rennes, France }
\affiliation {Laboratoire d'Annecy-le-Vieux de Physique des Particules (LAPP), Universit\'e de Savoie, CNRS/IN2P3, F-74941 Annecy-le-Vieux, France }
\affiliation {Washington State University, Pullman, WA 99164, USA }
\affiliation {Inter-University Centre for Astronomy and Astrophysics, Pune - 411007, India }
\affiliation {INFN, Sezione di Perugia, I-06123 Perugia, Italy }
\affiliation {Yukawa Institute for Theoretical Physics, Kyoto University, Kyoto 606-8502, Japan }
\affiliation {Universit\`a degli Studi di Urbino 'Carlo Bo', I-61029 Urbino, Italy }
\affiliation {INFN, Sezione di Firenze, I-50019 Sesto Fiorentino, Firenze, Italy }
\affiliation {University of Oregon, Eugene, OR 97403, USA }
\affiliation {Laboratoire Kastler Brossel, ENS, CNRS, UPMC, Universit\'e Pierre et Marie Curie, F-75005 Paris, France }
\affiliation {VU University Amsterdam, 1081 HV Amsterdam, The Netherlands }
\affiliation {University of Maryland, College Park, MD 20742, USA }
\affiliation {University of Massachusetts Amherst, Amherst, MA 01003, USA }
\affiliation {Laboratoire des Mat\'eriaux Avanc\'es (LMA), IN2P3/CNRS, Universit\'e de Lyon, F-69622 Villeurbanne, Lyon, France }
\affiliation {Universitat de les Illes Balears, E-07122 Palma de Mallorca, Spain }
\affiliation {Universit\`a di Napoli 'Federico II', Complesso Universitario di Monte S.Angelo, I-80126 Napoli, Italy }
\affiliation {Canadian Institute for Theoretical Astrophysics, University of Toronto, Toronto, Ontario, M5S 3H8, Canada }
\affiliation {Tsinghua University, Beijing 100084, China }
\affiliation {University of Michigan, Ann Arbor, MI 48109, USA }
\affiliation {INFN, Sezione di Roma Tor Vergata, I-00133 Roma, Italy }
\affiliation {National Tsing Hua University, Hsinchu Taiwan 300 }
\affiliation {Charles Sturt University, Wagga Wagga, NSW 2678, Australia }
\affiliation {Caltech-CaRT, Pasadena, CA 91125, USA }
\affiliation {Pusan National University, Busan 609-735, Korea }
\affiliation {Australian National University, Canberra, ACT 0200, Australia }
\affiliation {Carleton College, Northfield, MN 55057, USA }
\affiliation {Universit\`a di Roma Tor Vergata, I-00133 Roma, Italy }
\affiliation {INFN, Gran Sasso Science Institute, I-67100 L'Aquila, Italy }
\affiliation {Universit\`a di Roma 'La Sapienza', I-00185 Roma, Italy }
\affiliation {University of Brussels, Brussels 1050 Belgium }
\affiliation {Sonoma State University, Rohnert Park, CA 94928, USA }
\affiliation {The George Washington University, Washington, DC 20052, USA }
\affiliation {University of Cambridge, Cambridge, CB2 1TN, United Kingdom }
\affiliation {University of Minnesota, Minneapolis, MN 55455, USA }
\affiliation {The University of Texas at Brownsville, Brownsville, TX 78520, USA }
\affiliation {The University of Sheffield, Sheffield S10 2TN, United Kingdom }
\affiliation {Wigner RCP, RMKI, H-1121 Budapest, Konkoly Thege Mikl\'os \'ut 29-33, Hungary }
\affiliation {Argentinian Gravitational Wave Group, Cordoba Cordoba 5000, Argentina }
\affiliation {Universit\`a di Trento, I-38050 Povo, Trento, Italy }
\affiliation {INFN, Gruppo Collegato di Trento, I-38050 Povo, Trento, Italy }
\affiliation {Northwestern University, Evanston, IL 60208, USA }
\affiliation {Montclair State University, Montclair, NJ 07043, USA }
\affiliation {The Pennsylvania State University, University Park, PA 16802, USA }
\affiliation {MTA E\"otv\"os University, `Lendulet' A. R. G., Budapest 1117, Hungary }
\affiliation {Universit\`a di Perugia, I-06123 Perugia, Italy }
\affiliation {Rutherford Appleton Laboratory, HSIC, Chilton, Didcot, Oxon, OX11 0QX, United Kingdom }
\affiliation {Embry-Riddle Aeronautical University, Prescott, AZ 86301, USA }
\affiliation {Seoul National University, Seoul 151-742, Korea }
\affiliation {Perimeter Institute for Theoretical Physics, Waterloo, Ontario, N2L 2Y5, Canada }
\affiliation {American University, Washington, DC 20016, USA }
\affiliation {College of William and Mary, Williamsburg, VA 23187, USA }
\affiliation {University of Adelaide, Adelaide, SA 5005, Australia }
\affiliation {Raman Research Institute, Bangalore, Karnataka 560080, India }
\affiliation {Korea Institute of Science and Technology Information, Daejeon 305-806, Korea }
\affiliation {Bia{\l }ystok University, 15-424 Bia{\l }ystok, Poland }
\affiliation {University of Southampton, Southampton, SO17 1BJ, United Kingdom }
\affiliation {IISER-TVM, CET Campus, Trivandrum Kerala 695016, India }
\affiliation {Institute of Applied Physics, Nizhny Novgorod, 603950, Russia }
\affiliation {Hanyang University, Seoul 133-791, Korea }
\affiliation {NCBJ, 05-400 \'Swierk-Otwock, Poland }
\affiliation {IM-PAN, 00-956 Warsaw, Poland }
\affiliation {Institute for Plasma Research, Bhat, Gandhinagar 382428, India }
\affiliation {The University of Melbourne, Parkville, VIC 3010, Australia }
\affiliation {INFN, Sezione di Padova, I-35131 Padova, Italy }
\affiliation {Monash University, Victoria 3800, Australia }
\affiliation {SUPA, University of Strathclyde, Glasgow, G1 1XQ, United Kingdom }
\affiliation {Louisiana Tech University, Ruston, LA 71272, USA }
\affiliation {ESPCI, CNRS, F-75005 Paris, France }
\affiliation {Universit\`a di Camerino, Dipartimento di Fisica, I-62032 Camerino, Italy }
\affiliation {Southern University and A\&M College, Baton Rouge, LA 70813, USA }
\affiliation {IISER-Kolkata, Mohanpur, West Bengal 741252, India }
\affiliation {National Institute for Mathematical Sciences, Daejeon 305-390, Korea }
\affiliation {Hobart and William Smith Colleges, Geneva, NY 14456, USA }
\affiliation {RRCAT, Indore MP 452013, India }
\affiliation {Tata Institute for Fundamental Research, Mumbai 400005, India }
\affiliation {SUPA, University of the West of Scotland, Paisley, PA1 2BE, United Kingdom }
\affiliation {Institute of Astronomy, 65-265 Zielona G\'ora, Poland }
\affiliation {Indian Institute of Technology, Gandhinagar Ahmedabad Gujarat 382424, India }
\affiliation {Instituto de F\'\i sica Te\'orica, Univ. Estadual Paulista/ICTP South American Institute for Fundamental Research, S\~ao Paulo SP 01140-070, Brazil }
\affiliation {Andrews University, Berrien Springs, MI 49104, USA }
\affiliation {Trinity University, San Antonio, TX 78212, USA }
\affiliation {University of Washington, Seattle, WA 98195, USA }
\affiliation {Rochester Institute of Technology, Rochester, NY 14623, USA }
\affiliation {Southeastern Louisiana University, Hammond, LA 70402, USA }
\affiliation {Abilene Christian University, Abilene, TX 79699, USA }

\address{Ioffe Physical-Technical Institute, St. Petersburg, 194021, Russian Federation} 
\address{Universit\'{e} de Toulouse; UPS-OMP; IRAP; Toulouse, France} 
\address{CNRS; IRAP; 14, avenue Edouard Belin, F-31400 Toulouse, France} 
\address{CSPAR, University of Alabama in Huntsville, Huntsville, Alabama, USA} 
\address{University of California-Berkeley, Space Sciences Lab, 7 Gauss Way, Berkeley, CA 94720, USA} 
\address{Center for Research and Exploration in Space Science and Technology (CRESST) and NASA Goddard Space Flight Center, Greenbelt, MD 20771 USA} 
\address{Universities Space Research Association, 7178 Columbia Gateway Drive Columbia, MD 21046 USA} 
\address{INAF-IASF Bologna, Via P. Gobetti 101, 40129 Bologna, Italy} 
\address{St. Petersburg State Polytechnical University, 195251, St. Petersburg, Russia} 
\address{Los Alamos National Laboratory, Los Alamos, NM 87545 USA} 
\address{Graduate School of Science and Engineering, Saitama University, Saitama City Japan} 
\address{Max-Planck-Institut f\"{u}r extraterrestrische Physik, Giessenbachstrasse 1, 85748 Garching, Germany} 


%% file: L-V_ack_Jun2013.tex
The authors gratefully acknowledge the support of the United States
National Science Foundation for the construction and operation of the
LIGO Laboratory, the Science and Technology Facilities Council of the
United Kingdom, the Max-Planck-Society, and the State of
Niedersachsen/Germany for support of the construction and operation of
the GEO600 detector, and the Italian Istituto Nazionale di Fisica
Nucleare and the French Centre National de la Recherche Scientifique
for the construction and operation of the Virgo detector. The authors
also gratefully acknowledge the support of the research by these
agencies and by the Australian Research Council, 
the International Science Linkages program of the Commonwealth of Australia,
the Council of Scientific and Industrial Research of India, 
the Istituto Nazionale di Fisica Nucleare of Italy, 
the Spanish Ministerio de Econom\'ia y Competitividad,
the Conselleria d'Economia Hisenda i Innovaci\'o of the
Govern de les Illes Balears, the Foundation for Fundamental Research
on Matter supported by the Netherlands Organisation for Scientific Research, 
the Polish Ministry of Science and Higher Education, the FOCUS
Programme of Foundation for Polish Science,
the Royal Society, the Scottish Funding Council, the
Scottish Universities Physics Alliance, The National Aeronautics and
Space Administration, 
OTKA of Hungary,
the Lyon Institute of Origins (LIO),
the National Research Foundation of Korea,
Industry Canada and the Province of Ontario through the Ministry of Economic Development and Innovation, 
the National Science and Engineering Research Council Canada,
the Carnegie Trust, the Leverhulme Trust, the
David and Lucile Packard Foundation, the Research Corporation, and
the Alfred P. Sloan Foundation.